# Integrated Broadband Bowtie Antenna on Transparent Silica Substrate

Xingyu Zhang, *Member, IEEE*, Chi-Jui Chung, Shiyi Wang, Harish Subbaraman, *Member, IEEE*, Zeyu Pan, *Student Member, IEEE*, Qiwen Zhan, *Senior Member, IEEE*, and Ray T. Chen, *Fellow, IEEE*

*Abstract*— The bowtie antenna is a topic of growing interest in recent years. In this paper, we design, fabricate, and characterize a modified gold bowtie antenna integrated on a transparent silica substrate. The bowtie antenna is designed with broad RF bandwidth to cover the X-band in the electromagnetic spectrum. We numerically investigate the antenna characteristics, specifically its resonant frequency and enhancement factor. Our designed bowtie antenna provides a strong broadband electric field enhancement in its feed gap. Taking advantage of the low-k silica substrate, high enhancement factor can be achieved without the unwanted reflection and scattering from the backside silicon handle which is the issue of using an SOI substrate. We simulate the dependence of resonance frequency on bowtie geometry, and verify the simulation results through experimental investigation, by fabricating different sets of bowtie antennas on silica substrates and then measuring their resonance frequencies. In addition, the far-field radiation pattern of the bowtie antenna is measured, and it shows dipole-like characteristics with large beam width. Such a broadband antenna will be useful for a myriad of applications, ranging from photonic electromagnetic wave sensing to wireless communications.

## I. INTRODUCTION

In recent years, bowtie antenna has been a hot topic for intense theoretical and experimental investigation, impelled by its unique features and advantages, including simple planar structure, stable broadband performance, strong near-field enhancement [1-8]. One simple configuration used to achieve broadband characteristics is a biconical antenna formed by two cones, and the bowtie antenna is a simplified two-dimensional structure using two triangular shapes separated by a small gap that resembles the shape of a bowtie [9, 10]. The bowtie antenna concentrates the energy and provides a good localization of electric field inside the feed gap, providing a strong near-field enhancement [11, 12], which is useful for several applications, including optical sensing and energy harvesting [13, 14, 5, 15]. In addition, the resonant frequency of a bowtie antenna can be designed and tuned by appropriately modifying and scaling the bowtie geometry, such as its arm length, flare angle, and feed gap width. This enables the bowtie antenna to have various applications over a wide frequency range, including extreme-ultraviolet light generation [2], optical antennas [3, 16, 5], Terahertz-wave optoelectronics [17], mid-infrared plasmonic antennas and sensors [8, 18], microwave radar [7], wireless communications [4, 19], flexible RF devices [20], complex electromagnetic structures [21], and photonic detection of free-space electromagnetic waves [22-24].

For example, a modified gold bowtie antenna, consisting of a conventional bowtie shape with extension bars attached to its apex points, has been theoretically studied previously for microwave photonic applications [25]. Under RF illumination, an extended near-field area with a uniformly enhanced local electric field is obtained in its feed gap. Such an optimized bowtie antenna has recently been integrated with an electro-optic modulator inside its feed gap to form a very sensitive X-band electromagnetic wave sensor [26, 27]. For the detection of electromagnetic waves, the interaction of the waves and the substrate materials needs to be considered. In this case, a low-k dielectric substrate, such as a silica substrate (dielectric constant is about $\varepsilon_r$=3.9), is desired to provide better microwave coupling, because a low-k substrate would allow for higher received power (proportional to $1/\sqrt{\varepsilon_{eff}}$) [26]. More importantly, some electromagnetic wave sensors previously demonstrated on silicon-on-insulator (SOI) substrates [28, 27] suffer from unwanted reflection and scattering from backside silicon handle which has intrinsic doping [29], while, in comparison, the silica substrate (without backside silicon handle) can avoid this issue and improve their detection sensitivity. The investigation of the performance of the bowtie antenna on a silica substrate can provide a foundation for the next generation sensitive electromagnetic wave sensors made on silicon-on-glass substrates [30] or silicon-on-sapphire substrates [31][32].

In this work, we investigated the development and optimization of the abovementioned bowtie antenna on a cost-effective transparent low-k silica substrate. Numerical simulations on this type of bowtie antennas with different geometrical parameters are performed, and a series of measurements are carried out to verify the theoretical analysis. Specifically, we design, fabricate and experimentally characterize a broadband gold bowtie antenna on a silica substrate targeted around an operating frequency of 10.5GHz

This work was supported by Air Force Research Laboratory (AFRL) under the Small Business Technology Transfer Research (STTR) program (Grant No. FA8650-12-M-5131) monitored by Dr. Robert Nelson and Dr. Charles Lee.

X. Zhang is with Department of Electrical and Computer Engineering, University of Texas, Austin, TX 78758, USA (e-mail: xzhang@utexas.edu). He is now with Acacia Communications, Inc., Hazlet, NJ 07730, USA.

C.-J. Chung, Z. Pan, and R. T. Chen are with Department of Electrical and Computer Engineering, University of Texas, Austin, TX 78758, USA (e-mail: raychen@uts.cc.utexas.edu).

S. Wang and Q. Zhan are with Department of Electrical & Computer Engineering, University of Dayton, Dayton, OH 45469-2951, USA.

H. Subbaraman is with Omega Optics, Inc., Austin, TX 78757, USA.

(X-band). The designed bowtie antenna has a compact size smaller than 1cm². The electric-field enhancing capability of the bowtie antenna is investigated by numerical simulations. Its geometry-dependent resonant frequency is simulated, and then it is experimentally verified on a group of fabricated bowtie antennas on silica substrates. The radiation pattern of this bowtie antenna is also measured.

## II. DESIGN

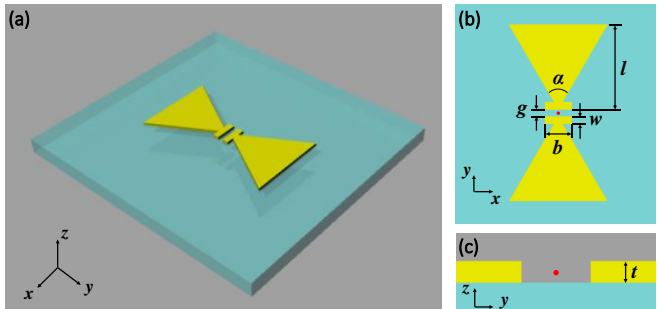

Figure 1. (a) 3D perspective of our modified gold bowtie antenna on a silica substrate. (b) Top view of the bowtie antenna. $l$: arm length; $\alpha$: flare angle; $g$: feed gap; $w$: bar width; $b$: bar length. (c) Cross section of the bowtie antenna. $t$: thickness. The red dot in the center of the feed gap at middle height indicates the observation point for the simulation of electric field enhancement.

A schematic of the modified bowtie antenna on a silica substrate is shown in Fig. 1, consisting of a conventional bowtie antenna with capacitive extension bars attached to the apex points of the bowtie [25]. The extension bars have a length, $b$=300μm, a width, $w$=10μm and a feed gap, $g$=10μm. The thickness of the gold film is chosen to be $t$=5μm, which is far beyond the skin depth of gold at the RF frequency of operation. The thickness of silica substrate is 1mm, which is not thick enough to adversely affect the broadside radiation of the X-band antenna [33]. Under RF illumination, highly enhanced local electric field is generated inside the feed gap of this bowtie antenna. The current on the bowtie antenna surface, induced by incident RF field, charges the feed gap and subsequently establishes this strong electric field in the feed gap [34]. Generally, the antenna system can be considered as a typical LC circuit, which is mainly composed of the inductive bowtie metallic arms and the capacitive bars, giving rise to an LC resonance determined by the antenna geometry. In this work, this resonance effect is characterized by field enhancement factor, which is defined as the resonant electric field amplitude at a specific observation point [red dot in Figs. 1 (b) and (c)] divided by the incident electric field amplitude. With the feed gap ($g$=10μm) and capacitive bars ($b$=300μm, and $w$=10μm) fixed, the resonant frequency of a bowtie antenna is mainly determined by the length of each bow arm and the flare angle $\alpha$ in Fig. 1 (b)) [35].

In order to show the uniform broadband electric field enhancement created over the entire feed gap, COMSOL Multiphysics is used to simulate a bowtie antenna model with an arm length of $l$=5.5mm and a flare angle of $\alpha$ =60°. The incident electric field is a continuous plane wave linearly polarized along the antenna axis (y-direction) and impinges upon the antenna from the top. Figure. 2 (a) shows the simulated field enhancement factor as a function of RF frequency. It can be seen that the bowtie antenna has a maximum field enhancement factor of ~688 at its resonant frequency of 10.5GHz, and that the 1-dB RF bandwidth is about 9GHz. Simulation results in Figs. 2 (b) and (c) show both the top view and the side view of the normalized local electric field amplitude at the resonant frequency. The electric field is mainly confined in the feed gap region, and it is strongly compressed by 688 times at the resonant frequency respect to the incident electric field. The polarization of the electric field is along the y direction in Figs. 2 (b) and (c). The peak resonant frequency can be tuned by adjusting the arm length and the flare angle of the bowtie antenna. The simulated resonant frequencies at different arm lengths and flare angles are shown in Fig. 5, respectively, and are correlated with experimental results in Section IV. More detailed theoretical analysis of the geometry-dependent performance of bowtie antennas can be found in Refs. [11, 36, 6].

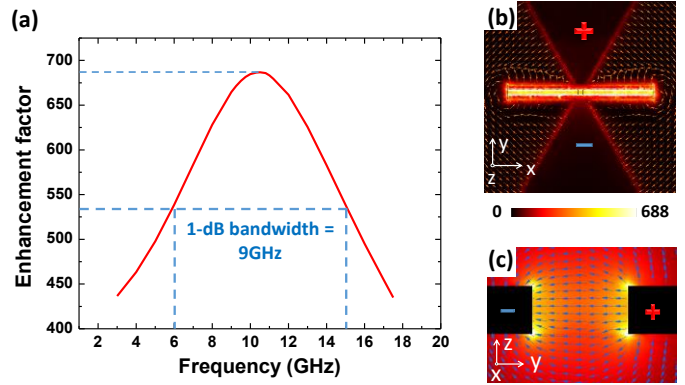

Figure 2. (a) The simulated field enhancement factor of the bowtie antenna with $l$=5.5mm and $\alpha$=60° as a function of RF frequencies. The maximum field enhancement factor is 688 at 10.5GHz, and the 1-dB RF bandwidth is 9GHz. (b) and (c) Top view and cross-sectional view of the simulated normalized electric field enhancement distribution at the resonant frequency.

## III. FABRICATION

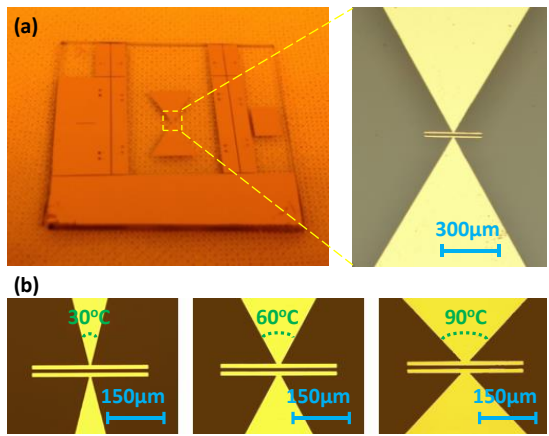

Figure 3. (a) A fabricated bowtie antenna on silica substrate. (b) Three bowtie antenna with flare angles of 30 degrees, 60 degrees and 90 degrees, respectively. For all these three bowtie antennas, the arm length $l$=4.5mm, the extended bar length $b$=300μm, extended bar width $w$=10μm, and the feed gap $g$=10μm.

Two groups of bowtie antennas are fabricated on 1mm-thick silica substrates ($\varepsilon_r$=~3.9) through standard CMOS manufacturing process. The first group of five bowtie antennas is fabricated with a fixed flare angle of 60 degrees, but arm lengths varying from 3.5mm to 5.5mm in steps of 0.5mm. The second group of three bowtie antennas has a fixed arm length of 4.5mm, but flare angles of 30 degrees, 60 degrees and 90 degrees, respectively.

First, a 50nm-thick gold seed layer with a 5nm-thick chromium adhesion buffer is deposited on the silica substrate by electron-beam evaporation. A 10 μm-thick AZ-9260 photoresist is spincoated on the seed layer, and it is then patterned as a buffer mask for the bowtie structure by photolithography. Next, a 5 μm-thick gold film is electroplated by through-mask plating method in a magnetically stirred electrolyte (Techni-Gold 25ES) at 50°C. Constant current of 8mA is applied through the solution in the entire electroplating process. After gold electroplating, the buffer mask is removed using Acetone, and then the gold/chromium seed layer is removed using wet etchants, leaving the electroplated gold bowtie structure on top of the transparent silica substrate. The conductivity of the electroplated gold film is measured to be $2.2 \times 10^7$ S/m. Microscope images of a few fabricated devices are shown in Fig. 3.

## IV. TESTING

First, in order to demonstrate the broadband characteristics of the fabricated bowtie antenna and to investigate the dependence of resonant frequency on the bowtie geometry, the two groups of bowtie antennas are tested as transmitting antennas. Each antenna is fed by applying a ground-signal (GS) microprobe (Cascade Microtech ACP40GS500) onto its two bow arms. A vector network analyzer (HP 8510C) is used to measure the transmission of these antennas over a broad frequency range of 1-20GHz. Assuming negligible loss, the normalized transmission signal can be inferred from the S11 measurements by zero minus S11 in dB scale. A broadband response can be clearly seen from the measured transmission signal, which agrees well with predication by simulation. The measured transmission signals of the first group of bowtie antennas are shown in Fig. 4 (a). The measured resonant frequencies as a function of arm lengths at a fixed flare angle of 60 degrees are extracted from the figure, and then correlated with the simulated resonant frequencies, as shown in Fig. 4 (b). Similarly, the measured resonant frequencies as a function of flare angles at a fixed arm length (e.g. 4.5mm) are extracted from the measured transmission signals of the second group of bowtie antennas, as shown in Fig. 4 (c), and then this measured results are correlated with simulated resonant frequencies in Fig. 4 (d). For longer bowtie arm or larger flare angle, the current flows through longer path to the gap, so the effective antenna size is increased, leading to longer resonant RF wavelength, which corresponds to lower resonant frequency. The trend of measured resonant frequencies agrees with the simulations. It can be seen that there are still some deviations between the measured and simulated resonant frequencies. This could be due to several reasons, such as the difference of the dielectric constant of an actual silica substrate and that assumed in simulations, and slight variations of size and shape of fabricated bowtie antenna from idealized model due to fabrication error.

Next, the far field radiation pattern is measured. The bowtie antenna with arm length of 5.5mm and flare angle of 60 degrees is used in this test. The bowtie antenna is mounted on a rotational stage and is rotated along the x axis (axes directions indicated in Fig. 1 or 2). RF signal from the vector network analyzer is coupled into the bowtie antenna through a GS microprobe. The frequency of this RF signal is set to 10.5GHz which is the resonant frequency of the bowtie antenna. An X-band horn antenna is placed 2m away as a receiving antenna in its far field region for the assumption of quasi-plane wave to hold. The received power is amplified by an RF amplifier and then measured by a microwave spectrum analyzer (HP 8560E). The measured normalized power as a function of rotation angle is shown as a blue curve in Fig. 5. Simulated radiation pattern (red curve) is also overlaid in the figure, showing a good match between simulation and experimental results. This measured radiation pattern indicates dipole-type characteristics of our bowtie antenna. The half power beam width is measured to be about 90 degrees. This wide beam width is good for the antenna to detect electromagnetic waves coming from a large range of incident angles.

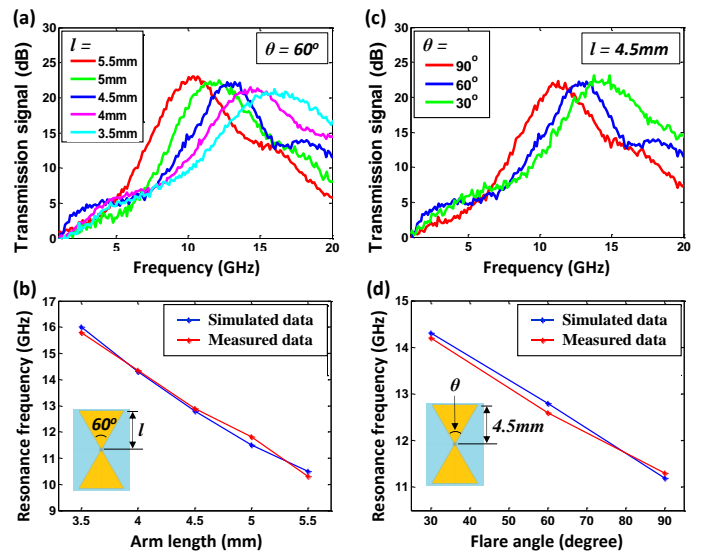

Figure 4. (a) Measured normalized transmission spectrum of bowtie antennas with different arm lengths. Flare angles are fixed at 60 degrees. (b) Correlation of measured resonant frequency with simulated resonant frequency at different arm lengths. (c) Measured normalized transmission spectrum of bowtie antennas with different flare angles. Arm lengths are fixed at 4.5mm. (d) Correlation of measured resonant frequency with simulated resonant frequency at different flare angles. $l$: arm length; $\alpha$: flare angle.

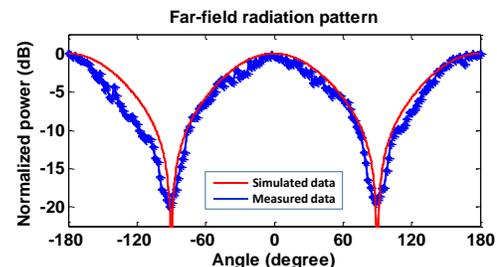

Figure 5. Measured far field radiation pattern (blue) of the bowtie antenna at a frequency of 10.5GHz. Simulated radiation pattern is also overlaid (red).

## V. CONCLUSION

In summary, we design, fabricate and experimentally demonstrate an integrated broadband bowtie antenna on a transparent silica substrate. Taking advantage of the low-k silica substrate, high enhancement factor can be achieved without the unwanted reflection and scattering from the backside silicon handle which is the issue of using an SOI substrate [28, 27]. The bowtie antenna is optimized to cover a broad frequency range in X-band. Numerical simulation shows that, with the arm length of 5.5mm and the flare angle of 60°, the electric field inside the bowtie feed gap can be enhanced by as high as 688 times at 10.5GHz compared to the incident electric field, with a 1-dB RF bandwidth over 9GHz. The dependency of resonant frequency on bowtie geometry, such as arm length and flare angle is numerically computed and experimentally verified. Based on the simulation and measurement results, in order to have a resonance frequency around 10.5GHz and broadly cover the center of X band, the geometry of the bowtie antenna can be optimized with the arm length of 5.5mm and the flare angle of 60°. For other different applications, the resonance frequency can also be tuned by appropriately modifying and scaling the bowtie geometry to cover other frequency ranges. In addition, the radiation pattern of the bowtie antenna is measured, showing a large angular beam width similar to a typical dipole antenna. The bowtie antenna has compact size smaller than 1cm². Compared with the theoretical analysis of the bowtie antenna on lithium niobate substrate in Ref. [25], our work presents an experimental demonstrate of this antenna on silica substrate.

With this bowtie antenna on silica substrate demonstrated, the electromagnetic field sensors on SOI substrates that we previously reported [27] can be transferred as silicon nanomembranes onto silica substrates [37] or directly fabricated on silicon-on-glass substrates or silicon-on-sapphire substrates [31, 30][38] to avoid impacts from backside silicon handles and thus to enhance their detection capability. Recently, a bowtie antenna coupled plasmonic electromagnetic wave sensor is theoretically proposed [39] and then experimentally demonstrated on a SOI substrate [40, 41]; however, the silicon substrate introduces high RF loss and weaken its electric field enhancement ability. Making the device on a silicon-on-glass substrate or a silicon-on-sapphire substrate and then using our optimized bowtie antenna in this work can provide a solution to address this problem and thus improve the sensitivity of the device. In addition to photonic electromagnetic wave sensing, this bowtie antenna has a various other potential applications, such as microwave radars, nano-antenna arrays, plasmonic sensing and Terahertz-wave detection [39]. Besides, this type of bowtie antennas can be fabricated by inkjet printing techniques in solid or contour forms [20, 42] on flexible substrates [20, 43], which is compatible with roll-to-roll manufacturing processes [44, 45]. Furthermore, the gold material can be replaced by ITO or graphene, together with the transparent feature of the silica substrate, to potentially enable some 'invisible' integrated electronic and photonic devices [46].

## REFERENCES

[1] R. C. Compton, R. C. McPhedran, Z. Popovic, G. M. Rebeiz, P. P. Tong, and D. B. Rutledge, "Bow-tie antennas on a dielectric half-space: theory and experiment," *Antennas and Propagation, IEEE Transactions on,* vol. 35, pp. 622-631, 1987.

[2] S. Kim, J. Jin, Y.-J. Kim, I.-Y. Park, Y. Kim, and S.-W. Kim, "High-harmonic generation by resonant plasmon field enhancement," *Nature,* vol. 453, pp. 757-760, 2008.

[3] K. D. Ko, A. Kumar, K. H. Fung, R. Ambekar, G. L. Liu, N. X. Fang, and K. C. Toussaint Jr, "Nonlinear optical response from arrays of Au bowtie nanoantennas," *Nano letters,* vol. 11, pp. 61-65, 2010.

[4] B. Madhav, V. Pisipati, H. Khan, V. Prasad, K. P. Kumar, K. Bhavani, and M. R. Kumar, "Liquid Crystal Bow-Tie Microstrip antenna for Wireless Communication Applications," *Journal of Engineering Science and Technology Review,* vol. 4, pp. 131-134, 2011.

[5] Z. Pan and J. Guo, "Enhanced optical absorption and electric field resonance in diabolo metal bar optical antennas," *Optics Express,* vol. 21, pp. 32491-32500, 2013.

[6] M. Rahim, M. Abdul Aziz, and C. Goh, "Bow-tie microstrip antenna design," in *13th IEEE International Conference,* 2005.

[7] M. Roslee, K. S. Subari, and I. S. Shahdan, "Design of bow tie antenna in CST studio suite below 2GHz for ground penetrating radar applications," in *RF and Microwave Conference (RFM), 2011 IEEE International,* 2011, pp. 430-433.

[8] S. Sederberg and A. Elezzabi, "Nanoscale plasmonic contour bowtie antenna operating in the mid-infrared," *Optics Express,* vol. 19, pp. 15532-15537, 2011.

[9] C. A. Balanis, *Antenna theory: analysis and design:* John Wiley & Sons, 2012.

[10] F. I. Rial, H. Lorenzo, M. Pereira, and J. Armesto, "Analysis of the emitted wavelet of high-resolution bowtie GPR Antennas," *Sensors,* vol. 9, pp. 4230-4246, 2009.

[11] H. Fischer and O. J. Martin, "Engineering the optical response of plasmonic nanoantennas," *Optics Express,* vol. 16, pp. 9144-9154, 2008.

[12] P. Mühlschlegel, H.-J. Eisler, O. Martin, B. Hecht, and D. Pohl, "Resonant optical antennas," *Science,* vol. 308, pp. 1607-1609, 2005.

[13] F. M. Congedo, G. Tarricone, and M. L. Cannarile, "Broadband bowtie antenna for RF energy scavenging applications," *Antennas and Propagation (EUCAP),* 2011.

[14] M. Mivelle, T. S. van Zanten, L. Neumann, N. F. van Hulst, and M. F. Garcia-Parajo, "Ultrabright bowtie nanoaperture antenna probes studied by single molecule fluorescence," *Nano letters,* vol. 12, pp. 5972-5978, 2012.

[15] J. Y. Suh, M. D. Huntington, C. H. Kim, W. Zhou, M. R. Wasielewski, and T. W. Odom, "Extraordinary nonlinear absorption in 3D bowtie nanoantennas," *Nano letters,* vol. 12, pp. 269-274, 2011.

[16] L. Novotny and N. Van Hulst, "Antennas for light," *Nature Photonics,* vol. 5, pp. 83-90, 2011.

[17] M. Jarrahi, "Advanced Photoconductive Terahertz Optoelectronics Based on Nano-Antennas and Nano-Plasmonic Light Concentrators," *IEEE Transactions on Terahertz Science and Technology,* vol. 5, pp. 391 - 397, 2015.

[18] W. Zhang, L. Huang, C. Santschi, and O. J. Martin, "Trapping and sensing 10 nm metal nanoparticles using plasmonic dipole antennas," *Nano letters,* vol. 10, pp. 1006-1011, 2010.

[19] N. Qi, Y. Xu, B. Chi, X. Yu, X. Zhang, N. Xu, P. Chiang, W. Rhee, and Z. Wang, "A dual-channel Compass/GPS/GLONASS/Galileo reconfigurable GNSS receiver in 65 nm CMOS with on-chip I/Q calibration," *Circuits and Systems I: Regular Papers, IEEE Transactions on,* vol. 59, pp. 1720-1732, 2012.

[20] A. C. Durgun, C. A. Balanis, C. R. Birtcher, and D. R. Allee, "Design, simulation, fabrication and testing of flexible bow-tie antennas," *Antennas and Propagation, IEEE Transactions on,* vol. 59, pp. 4425-4435, 2011.

[21] R. Chang, S. Li, M. Lubarda, B. Livshitz, and V. Lomakin, "FastMag: Fast micromagnetic simulator for complex magnetic structures," *Journal of Applied Physics,* vol. 109, p. 07D358, 2011.

[22] C.-Y. Lin, A. X. Wang, B. S. Lee, X. Zhang, and R. T. Chen, "High dynamic range electric field sensor for electromagnetic pulse detection," *Optics Express,* vol. 19, pp. 17372-17377, 2011.

[23] A. Savchenkov, W. Liang, V. Ilchenko, E. Dale, E. Savchenkova, A. Matsko, D. Seidel, and L. Maleki, "Photonic E-field sensor," *AIP Advances,* vol. 4, p. 122901, 2014.

[24] Y. N. Wijayanto, H. Murata, and Y. Okamura, "Electrooptic Millimeter-Wave–Lightwave Signal Converters Suspended to Gap-Embedded Patch Antennas on Low-Dielectric Materials," *Selected Topics in Quantum Electronics, IEEE Journal of,* vol. 19, pp. 3400709-3400709, 2013.

[25] S. Wang and Q. Zhan, "Modified bow-tie antenna with strong broadband field enhancement for RF photonic applications," in *SPIE NanoScience+ Engineering,* 2013, pp. 88061V-88061V-6.

[26] O. D. Herrera, K.-J. Kim, R. Voorakaranam, R. Himmelhuber, S. Wang, V. Demir, Q. Zhan, L. Li, R. A. Norwood, R. L. Nelson, J. Luo, A. K. Y. Jen, and N. Peyghambarian, "Silica/Electro-Optic Polymer Optical Modulator With Integrated Antenna for Microwave Receiving," *Journal of Lightwave Technology,* vol. 32, pp. 3861-3867, 2014/10/15 2014.

[27] X. Zhang, A. Hosseini, H. Subbaraman, S. Wang, Q. Zhan, J. Luo, A. K. Jen, and R. T. Chen, "Integrated Photonic Electromagnetic Field Sensor Based on Broadband Bowtie Antenna Coupled Silicon Organic Hybrid Modulator," *Lightwave Technology, Journal of,* vol. 32, pp. 3774-3784, 2014.

[28] L. Chen and R. M. Reano, "Compact electric field sensors based on indirect bonding of lithium niobate to silicon microrings," *Optics Express,* vol. 20, pp. 4032-4038, 2012.

[29] R. A. Synowicki, "Suppression of backside reflections from transparent substrates," *physica status solidi (c),* vol. 5, pp. 1085-1088, 2008.

[30] O. Nast, T. Puzzer, L. M. Koschier, A. B. Sproul, and S. R. Wenham, "Aluminum-induced crystallization of amorphous silicon on glass substrates above and below the eutectic temperature," *Applied Physics Letters,* vol. 73, pp. 3214-3216, 1998.

[31] R. Johnson, P. R. de la Houssaye, C. E. Chang, P.-F. Chen, M. E. Wood, G. A. Garcia, I. Lagnado, and P. M. Asbeck, "Advanced thin-film silicon-on-sapphire technology: microwave circuit applications," *Electron Devices, IEEE Transactions on,* vol. 45, pp. 1047-1054, 1998.

[32] C.-J. Chung, H. Subbaraman, J. Luo, A. K.-Y. Jen, R. L. Nelson, C. Y.-C. Lee, and R. T. Chen, "Fully packaged high-performance RF sensor featuring slotted photonic crystal waveguides in silicon-on-sapphire", SPIE Photonics West Conference, San Francisco, Feburary 18, 2016, Paper 9747-66 (Accepted)

[33] D. B. Rutledge, D. P. Neikirk, and D. P. Kasilingam, "Integrated circuit antennas," *Infrared and millimeter waves,* vol. 10, pp. 1-90, 1983.

[34] Q.-H. Park, "Optical antennas and plasmonics," *Contemporary Physics,* vol. 50, pp. 407-423, 2009.

[35] T. Søndergaard and S. Bozhevolnyi, "Slow-plasmon resonant nanostructures: Scattering and field enhancements," *Physical Review B,* vol. 75, p. 073402, 2007.

[36] J. George, M. Deepukumar, C. Aanandan, P. Mohanan, and K. Nair, "New compact microstrip antenna," *Electronics Letters,* vol. 32, pp. 508-509, 1996.

[37] J. Rogers, M. Lagally, and R. Nuzzo, "Synthesis, assembly and applications of semiconductor nanomembranes," *Nature,* vol. 477, pp. 45-53, 2011.

[38] C.-J. Chung, H. Subbaraman, J. Luo, A. K.-Y. Jen, R. L. Nelson, C. Y.-C. Lee, and R. T. Chen, "Fully packaged high-performance RF sensor featuring slotted photonic crystal waveguides in silicon-on-sapphire", SPIE Photonics West Conference, San Francisco, Feburary 18, 2016, Paper 9747-66 (Accepted)

[39] X. Zhang, A. Hosseini, H. Subbaraman, S. Wang, Q. Zhan, J. Luo, A. K. Jen, C.-j. Chung, H. Yan, and Z. Pan, "Antenna-coupled silicon-organic hybrid integrated photonic crystal modulator for broadband electromagnetic wave detection," in *SPIE OPTO*, 2015, pp. 93620O-93620O-20.

[40] Y. Salamin, R. Bonjour, W. Heni, C. Haffner, C. Hoessbacher, Y. Fedoryshyn, M. Zahner, D. Elder, L. Dalton, and C. Hafner, "Antenna Coupled Plasmonic Modulator," in *Frontiers in Optics*, 2015, p. FTh1B. 5.

[41] Y. Salamin, W. Heni, C. Haffner, Y. Fedoryshyn, C. Hoessbacher, R. Bonjour, M. Zahner, D. Hillerkuss, P. Leuchtmann, and D. L. Elder, "Direct Conversion of Free Space Millimeter Waves to Optical Domain by Plasmonic Modulator Antenna," *Nano letters,* 2015.

[42] A. A. Eldek, A. Z. Elsherbeni, and C. E. Smith, "Wideband microstrip-fed printed bow-tie antenna for phased-array systems," *Microwave and optical technology letters,* vol. 43, pp. 123-126, 2004.

[43] H. Subbaraman, D. T. Pham, X. C. Xu, M. Y. H. Chen, A. Hosseini, X. Lu, and R. T. Chen, "Inkjet-Printed Two-Dimensional Phased-Array Antenna on a Flexible Substrate," *IEEE Antennas and Wireless Propagation Letters,* vol. 12, pp. 170-173, 2013.

[44] X. Lin, T. Ling, H. Subbaraman, X. Zhang, K. Byun, L. J. Guo, and R. T. Chen, "Ultraviolet imprinting and aligned ink-jet printing for multilayer patterning of electro-optic polymer modulators," *Optics letters,* vol. 38, pp. 1597-1599, 2013.

[45] X. Lin, H. Subbaraman, Z. Pan, A. Hosseini, C. Longe, K. Kubena, P. Schleicher, P. Foster, S. Brickey, and R. T. Chen, "Towards Realizing High-Throughput, Roll-to-Roll Manufacturing of Flexible Electronic Systems," *Electronics,* vol. 3, pp. 624-635, 2014.

[46] A. Facchetti and T. J. Marks, "Transparent electronics," *From Synthesis to Applications, Wiley, Chichester, UK,* 2010.